\documentclass[aps,prl,twocolumn,superscriptaddress]{revtex4-1}
\usepackage{hyperref}
\usepackage{graphicx}
\usepackage{amsmath,braket}
\usepackage{upgreek}

\begin{document}

\title{Scalable squeezed light source for continuous variable quantum sampling}

\author{Z. Vernon}
\email{zach@xanadu.ai}
\affiliation{Xanadu, 372 Richmond St. W, Toronto, ON, M5V 1X6, Canada}
\author{N. Quesada}
\affiliation{Xanadu, 372 Richmond St. W, Toronto, ON, M5V 1X6, Canada}
\author{M. Liscidini}
\affiliation{Dipartimento di Fisica, Universit\`a degli studi di Pavia, Via Bassi 6, 27100 Pavia, Italy}
\affiliation{Impact Centre, University of Toronto, 411-112 College St., Toronto, ON, M5G 1L6, Canada} 
\author{B. Morrison}
\affiliation{Xanadu, 372 Richmond St. W, Toronto, ON, M5V 1X6, Canada}
\author{M. Menotti}
\affiliation{Xanadu, 372 Richmond St. W, Toronto, ON, M5V 1X6, Canada}
\author{K. Tan}
\affiliation{Xanadu, 372 Richmond St. W, Toronto, ON, M5V 1X6, Canada}
\author{J.E. Sipe}
\affiliation{Department of Physics, University of Toronto, 60 St. George St., Toronto, ON, M5S 1A7, Canada}


\date{\today}

\begin{abstract}
We propose a novel squeezed light source capable of meeting the stringent requirements of continuous variable quantum sampling. Using the effective $\chi_2$ interaction induced by a strong driving beam in the presence of the $\chi_3$ response in an integrated microresonator, our device is compatible with established nanophotonic fabrication platforms. With \emph{typical} realistic parameters, squeezed states with a mean photon number of 10 or higher can be generated in a single consistent temporal mode at repetition rates in excess of 100MHz. Over 15dB of squeezing is achievable in existing ultra-low loss platforms.
\end{abstract}

\pacs{}

\maketitle

Squeezed light is an essential resource for quantum information processing over continuous variables \cite{braunstein2005quantum}. Since the first measurements of small levels of squeezing were reported in the 1980s using hot atomic gases \cite{slusher1985observation} and then optical fiber \cite{shelby1986broad}, a number of techniques for its generation and control have been developed \cite{andersen201630}. Dominant among these techniques are those using parametric fluorescence in $\chi_2$ crystals \cite{wu1986generation}, and those exploiting the Kerr effect on short pulses in optical fibers \cite{silberhorn2001generation}. Both these techniques and others have enjoyed intensive development for achieving large squeezing levels \cite{vahlbruch2016detection}, low-frequency sideband squeezing \cite{mckenzie2004squeezing}, and tailoured spatiotemporal mode structure \cite{harder2016single}. These efforts have had a marked impact on squeezing-enhanced metrology \cite{caves1981quantum}, quantum computation \cite{lloyd1999quantum}, simulation \cite{huh2015boson}, and sampling \cite{hamilton2017gaussian}, as well as mesoscopic quantum optics \cite{harder2016single}.

Despite these efforts, to date no squeezed light source has been demonstrated that satisfies the many stringent requirements of full-scale continuous variable (CV) quantum computation and simulation \cite{weedbrook2012gaussian}. These are: (i) Scalability, i.e., the ease by which many tens or hundreds of identical mutually coherent and stabilized squeezed light sources can be integrated on one monolithic platform; (ii) Single-mode operation, i.e., the capability of producing squeezed light in a \emph{single} spatiotemporal mode, consistent across a wide range of squeezing levels, obviating the need for bulky and lossy mode-selective elements; (iii) Squeezing levels sufficient to enable a genuine quantum advantage in computation \cite{menicucci2014fault}, simulation \cite{huh2015boson}, and sampling \cite{hamilton2017gaussian}; (iv) Compatibility with single photon and photon number-resolving detection \cite{marsili2013detecting}, which are highly sensitive to noise from residual pump or spuriously generated light. The requirements (ii) and (iii) can be succinctly stated in mathematical terms: an ideal source provides an output quantum state of the form $e^{(r/2)a^2 - \mathrm{H.c.}}\vert\mathrm{vac}\rangle$, with squeezing factor $r$ reliably tunable, and in which $a$ is the annihilation operator for a single well-defined spatiotemporal mode, the characteristics of which do not vary over the tuning range of $r$.  In this work, we propose a scalable squeezed light source that comprehensively satisfies these requirements. We focus in particular on the application of quantum sampling in the Fock basis of large multimode Gaussian states. This application is a prime candidate for near-term demonstrations of quantum advantage, and can be used to address computational problems that truly are of practical interest \cite{huh2015boson,arrazola2018using,bradler2017gaussian}.  

Our device is based on the \emph{effective} $\chi_2$ interaction induced by a strong continuous wave (CW) coherent driving field in the presence of the $\chi_3$ response of an integrated nanophotonic resonator \cite{ramelow2018strong}. Combined with the resonance enhancement and tight transverse mode confinement provided by modern nanophotonic microresonators, this effective $\chi_2$ enables highly efficient parametric fluorescence when pumped by a secondary weaker field. Crucially, this device provides robust control over the spatiotemporal mode structure of the generated squeezed light. This allows the generation of true single-mode squeezed states in a benign, well-defined temporal mode at high repetition rates, consistent across a wide range of squeezing levels appropriate for CV quantum sampling.

\begin{figure}[!t]
\centering
\includegraphics[width=1.0\columnwidth]{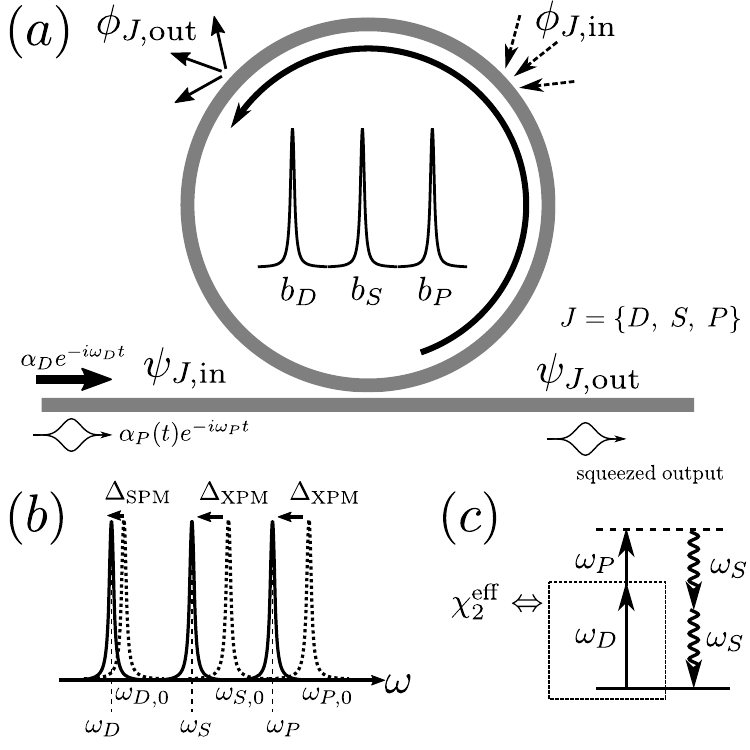}
\caption{(a) Microring resonator side-coupled to a channel waveguide. (b) Tuning the resonance condition for parametric fluorescence via self- and cross-phase modulation. (c) Virtual level diagram for dual-pumped spontaneous four-wave mixing; the strong CW drive beam mediates an effective second-order parametric nonlinearity $\chi_2^\mathrm{eff}$.}\label{fig:overview}
\end{figure}

An overview of the structure is illustrated in Fig.~\ref{fig:overview}a. We consider a high-finesse microresonator coupled to a channel waveguide; for definiteness we display a microring resonator geometry, though our theory and conclusions apply equally well to any microresonator. The ring accommodates a set of discrete resonant modes $J$, which we describe by annihilation operators $b_J$. In this work we focus on three key modes of interest: a drive mode $D$, signal mode $S$, and pump mode $P$. The intra-resonator Hamiltonian arising from the linear and third-order nonlinear response that connects these three modes is \cite{vernon2015spontaneous}
\begin{align}\label{eqn:hamiltonian}
H_\mathrm{res}=&\sum_J\hbar\omega_Jb_J^\dagger b_J -\hbar\Lambda\left(b_D b_P b_S^\dagger b_S^\dagger + \mathrm{H.c.}\right)  \\
-&\frac{\hbar\Lambda}{2}\sum_J b_J^\dagger b_J^\dagger b_J b_J \nonumber \\
-&2\hbar\Lambda \left(b_D^\dagger b_D(b_S^\dagger b_S + b_P^\dagger b_P) + b_P^\dagger b_P b_S^\dagger b_S\right)+ H_{X}, \nonumber
\end{align}
where $\omega_J$ is the resonant frequency of mode $J$, and $\Lambda$ is a constant related to the resonator geometry and third-order nonlinearity; for a microring, this is well approximated by $\Lambda \approx \hbar\omega_S v_g^2\gamma_{NL}/(2L)$ \cite{hoff2015integrated}, with $v_g$ the group velocity, $L$ the resonator length, and $\gamma_{NL}$ the waveguide nonlinear parameter. In general, such a resonator accommodates many more than the three modes of interest; the couplings of these extra modes to the $D$, $S$, and $P$ modes are contained in $H_X$, which we will examine shortly.

We restrict our analysis to the case in which the $D$ mode is driven by a strong coherent CW beam, yielding a large amplitude $\overline{\beta}_D e^{-i\omega_D t}$ in that mode, with $\overline{\beta}_D$ constant; for convenience we also take $\overline{\beta}_D$ to be real, which defines the phase reference for all other complex quantities. The first nonlinear term in (\ref{eqn:hamiltonian}) can then be written as $-\hbar\Lambda_2^{\mathrm{eff}}(t)b_Pb_S^\dagger b_S^\dagger + \mathrm{H.c.}$, in which $\Lambda_2^{\mathrm{eff}}(t)\equiv \Lambda\overline{\beta}_De^{-i\omega_D t}$. This situation is identical to that of a degenerate SPDC-like interaction driven by an effective second-order nonlinearity with strength $|\Lambda_2^\mathrm{eff}|$, which in this case has tunable magnitude determined by both the resonator's intrinsic nonlinearity, and the driving amplitude. 

As illustrated in Fig.~\ref{fig:overview}c, in the presence of this effective second-order nonlinearity a weaker coherent pump pulse in the $P$ mode thereby produces photon pairs via parametric fluorescence into the $S$ mode. This technique of using a strong CW pump in conjunction with the intrinsic $\chi_3$ response to mediate an effective $\chi_2$ interaction in an integrated microresonator has recently been demonstrated for the first time on a silicon nitride nanophotonic platform  to generate strong nonlinear mode coupling, giving rise to effective second harmonic generation with extremely large implemented conversion efficiency \cite{ramelow2018strong}. A similar dual pump scheme on a nanophotonic platform has also been used to drive optical parametric oscillation \cite{okawachi2015dual,reimer2015cross} and produce degenerate photon pairs \cite{he2015ultracompact,guo2014telecom}.
 
The second nonlinear term in (\ref{eqn:hamiltonian}) corresponds to self-phase modulation (SPM) of each mode, and the third to cross-phase modulation (XPM) between the three modes of interest. For the regime under consideration, in which the $D$ mode is driven by a strong CW beam, the $P$ mode by a much weaker CW or pulsed field, and in which the $S$ mode never accommodates a large mean photon number (i.e., well below any thresholds for parametric oscillation), we may neglect the effects of SPM and XPM due to photons in the $P$ mode and $S$ mode \cite{vernon2015spontaneous}. The effects of SPM and XPM are then completely encapsulated by static shifts in the effective resonance frequencies $\omega_J$ due to the large CW driving amplitude in the $D$ mode. The resonator Hamiltonian (\ref{eqn:hamiltonian}) under these circumstances can thus be well represented by
\begin{eqnarray}\label{eqn:hamiltonian_simplified}
H_\mathrm{res} &\to& \sum_J\hbar\omega_Jb_J^\dagger b_J - 2\hbar\Lambda |\beta_P(t)|^2b_S^\dagger b_S \nonumber \\
& & - \hbar|\Lambda_2^\mathrm{eff}|\left(\beta_P(t)b_S^\dagger b_S^\dagger + \mathrm{H.c.}\right) + H_X, 
\end{eqnarray}
where $\beta_P(t)=e^{-i(\omega_D+\omega_P)t}\overline{\beta}_P(t)$, with $\overline{\beta}_P(t)$ the (slowly varying) envelope of the intraresonator pulse amplitude in the $P$ mode, and in which the resonant frequencies $\omega_J$ are now understood to contain (drive power-dependent) corrections from the XPM-induced redshift due to the strong driving field. Though ultimately we will confine ourselves to a regime in which the pump amplitude $\beta_P(t)$ is sufficiently weak to have little effect on the $S$ mode from XPM, here we have retained the corresponding term to verify that fact in our calculations.

For the desired process to be phase-matched, in the simple ring system shown in Fig.~\ref{fig:overview}a the drive and pump resonances must be separated from the signal resonance by an equal number of mode orders; similarly, to maximize the efficiency of the process, the resonant frequencies of the three resonances must be close to evenly spaced. Absent any driving fields, material and modal dispersion in the resonator will give rise to a detuning $\Delta_\mathrm{res}=\omega_P+\omega_D-2\omega_S$ away from this  condition. As the driving beam power is increased to ``dress" the ring with an effective $\chi_2$, each resonant mode will experience a redshift in frequency due to SPM and XPM (in addition to a global thermal shift that is nearly the same for all three modes, and therefore may be neglected). Since the detuning $\Delta_\mathrm{XPM}$ from XPM is twice as large as the detuning $\Delta_\mathrm{SPM}$ from SPM, this effect can be used to counteract \emph{normal} dispersion: for a particular drive power and dispersion, the net detuning $\Delta_\mathrm{net}=\Delta_\mathrm{res} + \Delta_\mathrm{SPM} - \Delta_\mathrm{XPM}=\Delta_\mathrm{res} - \Delta_\mathrm{SPM}$ for the three modes can be reduced to zero. Thus, as illustrated in Fig.~\ref{fig:overview}b, the driving power can be used to tune the three phase-matched resonances into an equally spaced frequency configuration. 

A typical microresonator system accommodates many hundreds or even thousands of resonances. The term $H_X$ in (\ref{eqn:hamiltonian_simplified}) contains the corresponding contributions to the Hamiltonian, and their couplings to the three modes of interest. Below any thresholds for OPO and comb generation, and operating in a regime where cascaded four-wave mixing is negligible, there are two dominant unwanted couplings relevant to the device performance.  One gives rise to unwanted spontaneous four-wave mixing, leading to the generation of spurious photons in the $S$ mode; another gives rise to Bragg-scattering four-wave mixing, leading to an additional source of loss on the squeezed state generated in the $S$ mode \cite{vernon2016quantum,li2016efficient}. Both of these processes should be suppressed to yield a pure low-noise squeezed output. This can be accomplished by designing a system that suppresses the auxiliary resonances involved in their dynamics; many strategies have been demonstrated to selectively suppress certain resonances \cite{gentry2014tunable,chen2007compact}. More detail on these effects and strategies to eliminate them can be found in the Supplemenetal Information.

\begin{figure*}
\centering
\minipage{0.5\textwidth}
\centering
\includegraphics[width=1.0\linewidth,trim={0.0 0.0 0.0 0.2in},clip]{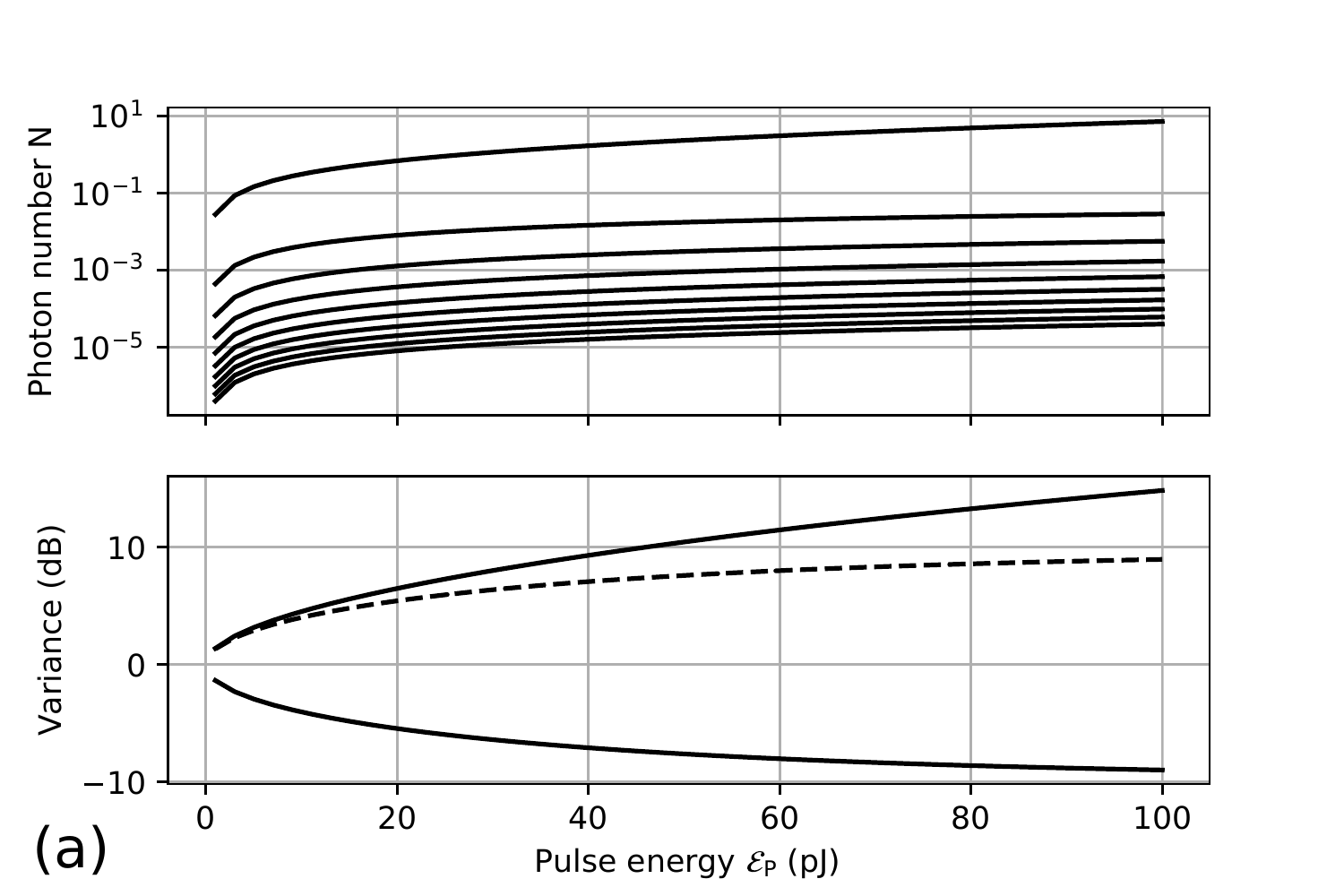}
\hfill\endminipage
\minipage{0.5\textwidth}
\centering
\includegraphics[width=1.0\linewidth]{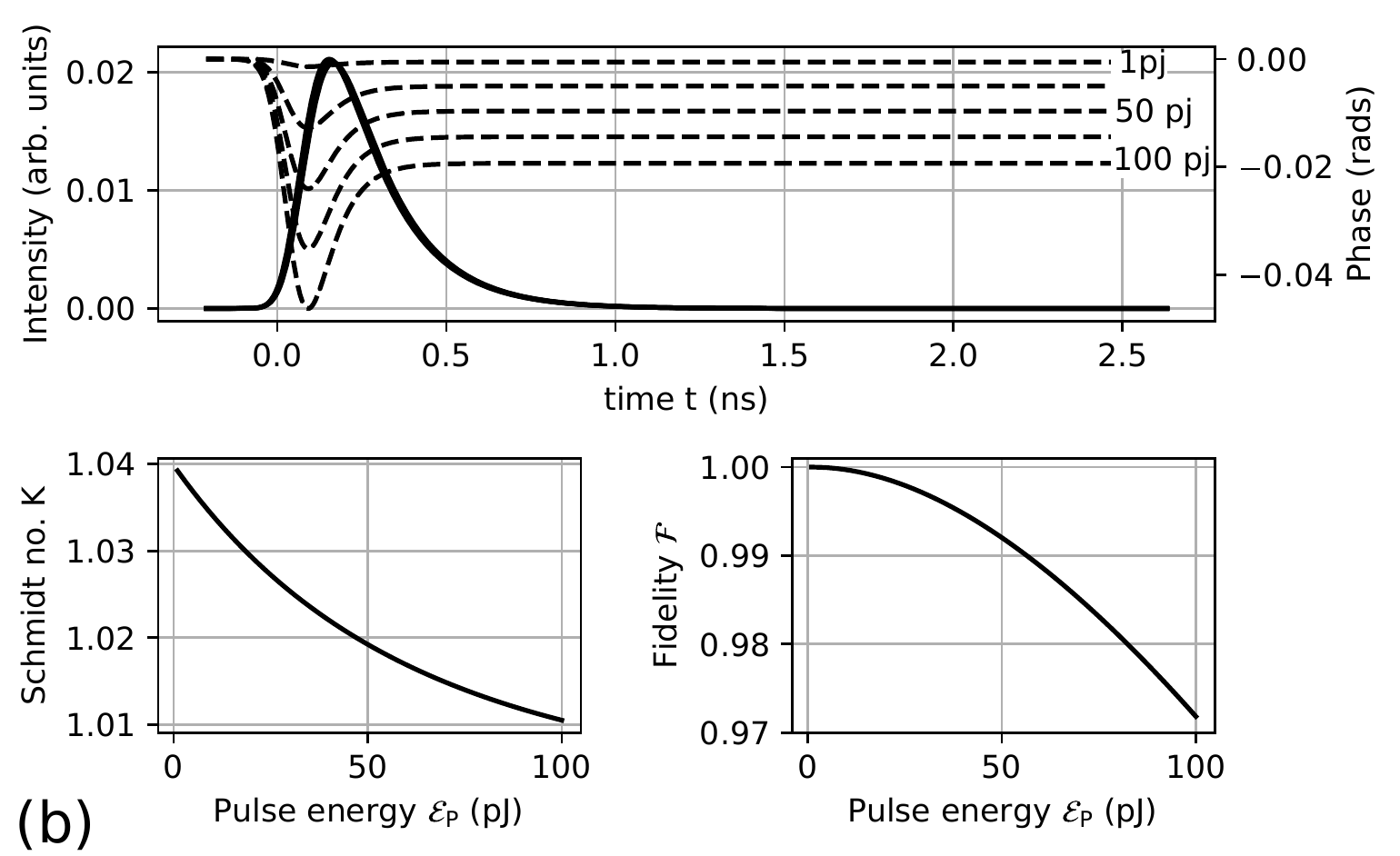}
\hfill\endminipage
\caption{System performance for a device with realistic parameters (details in text). (a) Top panel: Mean photon number of the first ten Schmidt modes as a function of pulse energy; the dominant mode (top curve) consistently lies about 100x above the next largest mode. Bottom panel: variance relative to vacuum of the squeezed quadrature (bottom solid curve) and anti-squeezed quadrature (top solid curve). Dashed curve shows variance of anti-squeezed quadrature for an ideal pure state; Some excess anti-squeezing is evident from the finite escape efficiency. (b) Top panel: Intensity (virtually identical solid curves) and phase (dashed curves) of temporal mode profile for the squeezed pulses generated for five pulse energies spanning 1 to 100 pJ. The intensity profile is virtually unchanged across this range; the phases show only very small progressive deviations due to cross-phase modulation from the pulsed pump as energy is increased, leading to very slight degradation of the fidelity between the complex pulse profile at each energy and that at the lowest energy (bottom right panel). The Schmidt number is consistently close to unity (bottom left panel). More details of how these quantities are extracted from the output moments can be found in the Supplemental Information. }\label{fig:squeezing_performance}
\end{figure*}

We now turn to calculating the properties of the squeezed output from the $S$ mode for a system appropriately designed to suppress unwanted processes, following a cavity input-output formalism appropriate for microresonators \cite{vernon2015spontaneous}. We consider a single-channel system like that shown in Fig \ref{fig:overview}a, for which we introduce Heisenberg-picture input and output field operators $\psi_{S,\mathrm{in}}(t)$  and $\psi_{S,\mathrm{out}}(t)$, as well as field operators $\phi_{S,\mathrm{in}}(t)$ and $\phi_{S,\mathrm{out}}(t)$ for the scattering modes that couple to the resonator modes due to the presence of loss. Here all time-dependent quantities are understood to be slowly varying, i.e., their fast optical dependence at $\omega_S$ has been removed; we also move into a rotating wave frame evolving as $e^{-i(\Delta_\mathrm{net}/2)t}$, taking into account a possible net detuning of the three resonances from the ideal evenly spaced configuration. The equation of motion for the resonator mode annihilation operator in this frame is then given by
\begin{eqnarray}\label{eqn:eqn_of_motion}
\frac{d}{dt}\begin{pmatrix}
b_S(t) \\
b_S^\dagger (t)
\end{pmatrix}
= M(t)\begin{pmatrix}
b_S(t) \\
b_S^\dagger (t)
\end{pmatrix}
+ \mathbf{d}_\mathrm{in}(t),
\end{eqnarray}
with the coupling matrix
\begin{eqnarray}
\lefteqn{M(t)=-\overline{\Gamma}_SI_2} \\
&& + \begin{pmatrix}
  i\left(\frac{\Delta_\mathrm{net}}{2} + 2\Lambda|\overline{\beta}_P(t)|^2\right) & g(t) \\
g^*(t) &  - i\left(\frac{\Delta_\mathrm{net}}{2} + 2\Lambda|\overline{\beta}_P(t)|^2\right)
\end{pmatrix}\nonumber,
\end{eqnarray}
with $I_2$ the $2\times 2$ identity matrix,
and input vacuum fluctuations $\mathbf{d}_\mathrm{in}(t)=\left(-i\gamma_S^*\psi_{S,\mathrm{in}}(t) - i\mu_S^*\phi_{S,\mathrm{in}}(t),i\gamma_S\psi_{S,\mathrm{in}}^\dagger(t) + i\mu_S\phi_{S,\mathrm{in}}^\dagger(t)\right)^T$. Here the function $g(t)\equiv 2i\Lambda\overline{\beta}_D\overline{\beta}_P(t)$ describes the time-dependent pump and nonlinear strength in the resonator, and $\overline{\Gamma}_S=\Gamma_S+M_S$ the total damping rate of the resonator $S$ mode, to which both scattering loss (with associated rate $M_S=|\mu_S|^2/(2v_g)$) and the resonator-channel coupling (with associated rate $\Gamma_S=|\gamma_S|^2/(2v_g)$) contribute. These are related to the total loaded quality factor $Q_S$ of the $S$ resonance via $Q_S=\omega_S/(2\overline{\Gamma}_S)$; the escape efficiency for that resonance is then given by $\eta^\mathrm{esc}_S=\Gamma_S/\overline{\Gamma}_S$.

The output field in the channel is given by $\psi_{S,\mathrm{out}}(t)=\psi_{S,\mathrm{in}}(t) - i(\gamma_S/v_g)b_S(t)$. Thus a solution for $b_S(t)$ enables the calculation of all properties of the output. For the linearized dynamics of (\ref{eqn:eqn_of_motion}), it is straightforward to construct a Green function for the system response: a solution is given by $(b_S(t),b_S^\dagger(t))^T=\int_{-\infty}^t dt' G(t,t')\mathbf{d}_\mathrm{in}(t')$, where the $2\times 2$ matrix Green function $G$ satisfies $G(t,t)=I_2$ for all $t$, and $dG(t,t')/dt=M(t)G(t,t')$ for $t>t'$. This equation can be solved numerically; the properties of all system outputs can then be expressed in terms of the four components of $G(t,t')$. In addition to the static system parameters (quality factor, coupling ratios, etc.), the function $g(t)$ also must be specified; this can be calculated by numerically integrating the associated nonlinear equation of motion for the pulsed mode, given by $d\overline{\beta}_P(t)/dt=(-\overline{\Gamma}_P + i\Lambda |\overline{\beta}_P(t)|^2)\overline{\beta}_P(t) - i\gamma_P^*\alpha_{P,\mathrm{in}}(t)$, where $\alpha_{P,\mathrm{in}}$ is the input pump pulse profile in the channel, normalized such that the energy $\mathcal{E}_P$ in the pulse is given by $\mathcal{E}_P=\hbar\omega_Pv_g\int_{-\infty}^\infty dt |\alpha_{P,\mathrm{in}}(t)|^2$. With the drive mode amplitude $\overline{\beta}_D$ and the nonlinear strength $\Lambda$, $g(t)$ is determined; for a resonant drive beam, $\overline{\beta}_D$ is given by $\overline{\beta}_D=2\sqrt{P_D Q_D \eta^\mathrm{esc}_D/(\hbar\omega_D^2)}$, with $P_D$ the input drive power in the channel, $Q_D$ the full loaded quality factor of the drive resonance, and $\eta^\mathrm{esc}_D$ the associated escape efficiency.

Since the dynamics of the system are linear in the mode operators, for a vacuum input in the $S$ mode the output must correspond to a Gaussian state with zero mean. Thus all properties of the system output can be expressed in terms of the second-order moments $N(t,t')=v_g\langle \psi_{S,\mathrm{out}}^\dagger(t)\psi_{S,\mathrm{out}}(t')\rangle$ and $M(t,t')=v_g\langle \psi_{S,\mathrm{out}}(t)\psi_{S,\mathrm{out}}(t')\rangle$. From these moments we can extract all measurable quantities. For our purposes we are primarily concerned with those aspects of the system output that are relevant for applications in CV quantum sampling: the efficiency (degree of squeezing as a function of drive and pump pulse powers), purity (limited by scattering losses), and temporal mode structure. The latter can be assessed by calculating the Schmidt number $K$ of the output field, which quantifies the number of excited output modes, and is ideally close to unity for single-temporal-mode squeezed states. It is also important to assess the full complex temporal mode shape of the generated squeezed light, to ensure that is benign (i.e., does not suffer from a complicated and erratic phase structure or envelope), and consistent across a wide range of squeezing levels. This last point is \emph{crucial} for CV quantum sampling applications, for which squeezed states with different squeezing levels must interfere.

We examine a system with realistic device parameters, well below best-reported values, that are routinely achievable in modern silicon nitride microring resonators \cite{moss2013new}. We consider a device optimized for a CW drive input power of 200 mW at the phase-matching point that yields $\Delta_\mathrm{net}=0$, and include the effects of time-dependent self-phase modulation and cross-phase modulation from the pump pulse. In Fig.~ \ref{fig:squeezing_performance} the squeezing performance is illustrated for such a device with $400\;\upmu\mathrm{m}$ round-trip length length, $\omega_S=2\pi\times 193$ THz, nonlinear parameter $\gamma_\mathrm{NL}=1\;(\mathrm{Wm})^{-1}$, group velocity $v_g=c/1.7$, and intrinsic quality factor of $2\times 10^6$ for all three resonances with escape efficiencies of $0.5$ (critically coupled) for the drive mode $D$, $0.9$ for the $S$ mode, and $0.98$ for the pump mode $P$; the corresponding loaded quality factors are then respectively $1\times10^6$, $2\times 10^5$, and $4\times 10^4$. This sequence of coupling ratios is chosen for maximal circulating power in the $D$ mode, good escape efficiency for the generated photons in the $S$ mode, and to allow large-bandwidth pulses into the $P$ mode, which is necessary for achieving low Schmidt number \cite{vernon2017truly,christensen2018engineering}. Independent control over the escape efficiencies can be realized by suitable coupler design; many strategies exist to accomplish this \cite{tison2017path}. 

The fundamental limit to squeezing attainable in this system is set by the escape efficiency, which in this case limits the output to $-10\log(1-\eta_S^\mathrm{esc}) = 10$ dB of squeezing. As evident from Fig.~\ref{fig:squeezing_performance}a, the system can readily approach loss-limited performance, with nearly 10 dB of squeezing realized for a Gaussian pump pulse having energy 100 pJ and intensity full width at half maximum duration set to one tenth of the $S$ mode dwelling time. This level of squeezing is precisely the desired operational point for many CV quantum sampling protocols, which typically call for squeezed states with a mean photon number of about one \cite{hamilton2017gaussian}. Furthermore, as shown in Fig.~\ref{fig:squeezing_performance}b, the system produces clean single temporal-mode squeezed pulses of roughly 1ns duration, with negligible variation in their pulse profiles across a wide tuning range of squeezing levels. The Schmidt number, and the fidelity of the generated temporal mode at high input energies with that at low input energies, both remain very close to unity. For applications requiring very high squeezing levels, such as metrology and CV teleportation \cite{menicucci2014fault,braunstein1998teleportation}, we note that existing ultra-low loss platforms \cite{spencer2014integrated} permit the signal resonance escape efficiency to be further optimized while maintaining acceptable efficiency; for a system with intrinsic quality factors of $10^7$ with $\eta_S^\mathrm{esc}=0.99$, $\eta_P^\mathrm{esc}=0.999$, and otherwise identical parameters, 15dB of squeezing is realized with only a few dB of excess anti-squeezing arising from the sub-unity escape efficiency. We therefore expect this proposed strategy to be of considerable utility for a wide range of CV quantum information processing applications.

\begin{acknowledgments}
The authors acknowledge support from the Ontario Centres of Excellence.
\end{acknowledgments}

\bibliography{bib-dualpump}

\begin{thebibliography}{40}%
\makeatletter
\providecommand \@ifxundefined [1]{%
 \@ifx{#1\undefined}
}%
\providecommand \@ifnum [1]{%
 \ifnum #1\expandafter \@firstoftwo
 \else \expandafter \@secondoftwo
 \fi
}%
\providecommand \@ifx [1]{%
 \ifx #1\expandafter \@firstoftwo
 \else \expandafter \@secondoftwo
 \fi
}%
\providecommand \natexlab [1]{#1}%
\providecommand \enquote  [1]{``#1''}%
\providecommand \bibnamefont  [1]{#1}%
\providecommand \bibfnamefont [1]{#1}%
\providecommand \citenamefont [1]{#1}%
\providecommand \href@noop [0]{\@secondoftwo}%
\providecommand \href [0]{\begingroup \@sanitize@url \@href}%
\providecommand \@href[1]{\@@startlink{#1}\@@href}%
\providecommand \@@href[1]{\endgroup#1\@@endlink}%
\providecommand \@sanitize@url [0]{\catcode `\\12\catcode `\$12\catcode
  `\&12\catcode `\#12\catcode `\^12\catcode `\_12\catcode `\%12\relax}%
\providecommand \@@startlink[1]{}%
\providecommand \@@endlink[0]{}%
\providecommand \url  [0]{\begingroup\@sanitize@url \@url }%
\providecommand \@url [1]{\endgroup\@href {#1}{\urlprefix }}%
\providecommand \urlprefix  [0]{URL }%
\providecommand \Eprint [0]{\href }%
\providecommand \doibase [0]{http://dx.doi.org/}%
\providecommand \selectlanguage [0]{\@gobble}%
\providecommand \bibinfo  [0]{\@secondoftwo}%
\providecommand \bibfield  [0]{\@secondoftwo}%
\providecommand \translation [1]{[#1]}%
\providecommand \BibitemOpen [0]{}%
\providecommand \bibitemStop [0]{}%
\providecommand \bibitemNoStop [0]{.\EOS\space}%
\providecommand \EOS [0]{\spacefactor3000\relax}%
\providecommand \BibitemShut  [1]{\csname bibitem#1\endcsname}%
\let\auto@bib@innerbib\@empty
\bibitem [{\citenamefont {Braunstein}\ and\ \citenamefont
  {Van~Loock}(2005)}]{braunstein2005quantum}%
  \BibitemOpen
  \bibfield  {author} {\bibinfo {author} {\bibfnamefont {S.~L.}\ \bibnamefont
  {Braunstein}}\ and\ \bibinfo {author} {\bibfnamefont {P.}~\bibnamefont
  {Van~Loock}},\ }\href@noop {} {\bibfield  {journal} {\bibinfo  {journal}
  {Reviews of Modern Physics}\ }\textbf {\bibinfo {volume} {77}},\ \bibinfo
  {pages} {513} (\bibinfo {year} {2005})}\BibitemShut {NoStop}%
\bibitem [{\citenamefont {Slusher}\ \emph {et~al.}(1985)\citenamefont
  {Slusher}, \citenamefont {Hollberg}, \citenamefont {Yurke}, \citenamefont
  {Mertz},\ and\ \citenamefont {Valley}}]{slusher1985observation}%
  \BibitemOpen
  \bibfield  {author} {\bibinfo {author} {\bibfnamefont {R.}~\bibnamefont
  {Slusher}}, \bibinfo {author} {\bibfnamefont {L.}~\bibnamefont {Hollberg}},
  \bibinfo {author} {\bibfnamefont {B.}~\bibnamefont {Yurke}}, \bibinfo
  {author} {\bibfnamefont {J.}~\bibnamefont {Mertz}}, \ and\ \bibinfo {author}
  {\bibfnamefont {J.}~\bibnamefont {Valley}},\ }\href@noop {} {\bibfield
  {journal} {\bibinfo  {journal} {Physical Review Letters}\ }\textbf {\bibinfo
  {volume} {55}},\ \bibinfo {pages} {2409} (\bibinfo {year}
  {1985})}\BibitemShut {NoStop}%
\bibitem [{\citenamefont {Shelby}\ \emph {et~al.}(1986)\citenamefont {Shelby},
  \citenamefont {Levenson}, \citenamefont {Perlmutter}, \citenamefont {DeVoe},\
  and\ \citenamefont {Walls}}]{shelby1986broad}%
  \BibitemOpen
  \bibfield  {author} {\bibinfo {author} {\bibfnamefont {R.}~\bibnamefont
  {Shelby}}, \bibinfo {author} {\bibfnamefont {M.}~\bibnamefont {Levenson}},
  \bibinfo {author} {\bibfnamefont {S.}~\bibnamefont {Perlmutter}}, \bibinfo
  {author} {\bibfnamefont {R.}~\bibnamefont {DeVoe}}, \ and\ \bibinfo {author}
  {\bibfnamefont {D.}~\bibnamefont {Walls}},\ }\href@noop {} {\bibfield
  {journal} {\bibinfo  {journal} {Physical Review Letters}\ }\textbf {\bibinfo
  {volume} {57}},\ \bibinfo {pages} {691} (\bibinfo {year} {1986})}\BibitemShut
  {NoStop}%
\bibitem [{\citenamefont {Andersen}\ \emph {et~al.}(2016)\citenamefont
  {Andersen}, \citenamefont {Gehring}, \citenamefont {Marquardt},\ and\
  \citenamefont {Leuchs}}]{andersen201630}%
  \BibitemOpen
  \bibfield  {author} {\bibinfo {author} {\bibfnamefont {U.~L.}\ \bibnamefont
  {Andersen}}, \bibinfo {author} {\bibfnamefont {T.}~\bibnamefont {Gehring}},
  \bibinfo {author} {\bibfnamefont {C.}~\bibnamefont {Marquardt}}, \ and\
  \bibinfo {author} {\bibfnamefont {G.}~\bibnamefont {Leuchs}},\ }\href@noop {}
  {\bibfield  {journal} {\bibinfo  {journal} {Physica Scripta}\ }\textbf
  {\bibinfo {volume} {91}},\ \bibinfo {pages} {053001} (\bibinfo {year}
  {2016})}\BibitemShut {NoStop}%
\bibitem [{\citenamefont {Wu}\ \emph {et~al.}(1986)\citenamefont {Wu},
  \citenamefont {Kimble}, \citenamefont {Hall},\ and\ \citenamefont
  {Wu}}]{wu1986generation}%
  \BibitemOpen
  \bibfield  {author} {\bibinfo {author} {\bibfnamefont {L.-A.}\ \bibnamefont
  {Wu}}, \bibinfo {author} {\bibfnamefont {H.}~\bibnamefont {Kimble}}, \bibinfo
  {author} {\bibfnamefont {J.}~\bibnamefont {Hall}}, \ and\ \bibinfo {author}
  {\bibfnamefont {H.}~\bibnamefont {Wu}},\ }\href@noop {} {\bibfield  {journal}
  {\bibinfo  {journal} {Physical Review Letters}\ }\textbf {\bibinfo {volume}
  {57}},\ \bibinfo {pages} {2520} (\bibinfo {year} {1986})}\BibitemShut
  {NoStop}%
\bibitem [{\citenamefont {Silberhorn}\ \emph {et~al.}(2001)\citenamefont
  {Silberhorn}, \citenamefont {Lam}, \citenamefont {Weiss}, \citenamefont
  {K{\"o}nig}, \citenamefont {Korolkova},\ and\ \citenamefont
  {Leuchs}}]{silberhorn2001generation}%
  \BibitemOpen
  \bibfield  {author} {\bibinfo {author} {\bibfnamefont {C.}~\bibnamefont
  {Silberhorn}}, \bibinfo {author} {\bibfnamefont {P.~K.}\ \bibnamefont {Lam}},
  \bibinfo {author} {\bibfnamefont {O.}~\bibnamefont {Weiss}}, \bibinfo
  {author} {\bibfnamefont {F.}~\bibnamefont {K{\"o}nig}}, \bibinfo {author}
  {\bibfnamefont {N.}~\bibnamefont {Korolkova}}, \ and\ \bibinfo {author}
  {\bibfnamefont {G.}~\bibnamefont {Leuchs}},\ }\href@noop {} {\bibfield
  {journal} {\bibinfo  {journal} {Physical Review Letters}\ }\textbf {\bibinfo
  {volume} {86}},\ \bibinfo {pages} {4267} (\bibinfo {year}
  {2001})}\BibitemShut {NoStop}%
\bibitem [{\citenamefont {Vahlbruch}\ \emph {et~al.}(2016)\citenamefont
  {Vahlbruch}, \citenamefont {Mehmet}, \citenamefont {Danzmann},\ and\
  \citenamefont {Schnabel}}]{vahlbruch2016detection}%
  \BibitemOpen
  \bibfield  {author} {\bibinfo {author} {\bibfnamefont {H.}~\bibnamefont
  {Vahlbruch}}, \bibinfo {author} {\bibfnamefont {M.}~\bibnamefont {Mehmet}},
  \bibinfo {author} {\bibfnamefont {K.}~\bibnamefont {Danzmann}}, \ and\
  \bibinfo {author} {\bibfnamefont {R.}~\bibnamefont {Schnabel}},\ }\href@noop
  {} {\bibfield  {journal} {\bibinfo  {journal} {Physical Review Letters}\
  }\textbf {\bibinfo {volume} {117}},\ \bibinfo {pages} {110801} (\bibinfo
  {year} {2016})}\BibitemShut {NoStop}%
\bibitem [{\citenamefont {McKenzie}\ \emph {et~al.}(2004)\citenamefont
  {McKenzie}, \citenamefont {Grosse}, \citenamefont {Bowen}, \citenamefont
  {Whitcomb}, \citenamefont {Gray}, \citenamefont {McClelland},\ and\
  \citenamefont {Lam}}]{mckenzie2004squeezing}%
  \BibitemOpen
  \bibfield  {author} {\bibinfo {author} {\bibfnamefont {K.}~\bibnamefont
  {McKenzie}}, \bibinfo {author} {\bibfnamefont {N.}~\bibnamefont {Grosse}},
  \bibinfo {author} {\bibfnamefont {W.~P.}\ \bibnamefont {Bowen}}, \bibinfo
  {author} {\bibfnamefont {S.~E.}\ \bibnamefont {Whitcomb}}, \bibinfo {author}
  {\bibfnamefont {M.~B.}\ \bibnamefont {Gray}}, \bibinfo {author}
  {\bibfnamefont {D.~E.}\ \bibnamefont {McClelland}}, \ and\ \bibinfo {author}
  {\bibfnamefont {P.~K.}\ \bibnamefont {Lam}},\ }\href@noop {} {\bibfield
  {journal} {\bibinfo  {journal} {Physical Review Letters}\ }\textbf {\bibinfo
  {volume} {93}},\ \bibinfo {pages} {161105} (\bibinfo {year}
  {2004})}\BibitemShut {NoStop}%
\bibitem [{\citenamefont {Harder}\ \emph {et~al.}(2016)\citenamefont {Harder},
  \citenamefont {Bartley}, \citenamefont {Lita}, \citenamefont {Nam},
  \citenamefont {Gerrits},\ and\ \citenamefont
  {Silberhorn}}]{harder2016single}%
  \BibitemOpen
  \bibfield  {author} {\bibinfo {author} {\bibfnamefont {G.}~\bibnamefont
  {Harder}}, \bibinfo {author} {\bibfnamefont {T.~J.}\ \bibnamefont {Bartley}},
  \bibinfo {author} {\bibfnamefont {A.~E.}\ \bibnamefont {Lita}}, \bibinfo
  {author} {\bibfnamefont {S.~W.}\ \bibnamefont {Nam}}, \bibinfo {author}
  {\bibfnamefont {T.}~\bibnamefont {Gerrits}}, \ and\ \bibinfo {author}
  {\bibfnamefont {C.}~\bibnamefont {Silberhorn}},\ }\href@noop {} {\bibfield
  {journal} {\bibinfo  {journal} {Physical Review Letters}\ }\textbf {\bibinfo
  {volume} {116}},\ \bibinfo {pages} {143601} (\bibinfo {year}
  {2016})}\BibitemShut {NoStop}%
\bibitem [{\citenamefont {Caves}(1981)}]{caves1981quantum}%
  \BibitemOpen
  \bibfield  {author} {\bibinfo {author} {\bibfnamefont {C.~M.}\ \bibnamefont
  {Caves}},\ }\href@noop {} {\bibfield  {journal} {\bibinfo  {journal}
  {Physical Review D}\ }\textbf {\bibinfo {volume} {23}},\ \bibinfo {pages}
  {1693} (\bibinfo {year} {1981})}\BibitemShut {NoStop}%
\bibitem [{\citenamefont {Lloyd}\ and\ \citenamefont
  {Braunstein}(1999)}]{lloyd1999quantum}%
  \BibitemOpen
  \bibfield  {author} {\bibinfo {author} {\bibfnamefont {S.}~\bibnamefont
  {Lloyd}}\ and\ \bibinfo {author} {\bibfnamefont {S.~L.}\ \bibnamefont
  {Braunstein}},\ }\href@noop {} {\bibfield  {journal} {\bibinfo  {journal}
  {Physical Review Letters}\ }\textbf {\bibinfo {volume} {82}},\ \bibinfo
  {pages} {1784} (\bibinfo {year} {1999})}\BibitemShut {NoStop}%
\bibitem [{\citenamefont {Huh}\ \emph {et~al.}(2015)\citenamefont {Huh},
  \citenamefont {Guerreschi}, \citenamefont {Peropadre}, \citenamefont
  {McClean},\ and\ \citenamefont {Aspuru-Guzik}}]{huh2015boson}%
  \BibitemOpen
  \bibfield  {author} {\bibinfo {author} {\bibfnamefont {J.}~\bibnamefont
  {Huh}}, \bibinfo {author} {\bibfnamefont {G.~G.}\ \bibnamefont {Guerreschi}},
  \bibinfo {author} {\bibfnamefont {B.}~\bibnamefont {Peropadre}}, \bibinfo
  {author} {\bibfnamefont {J.~R.}\ \bibnamefont {McClean}}, \ and\ \bibinfo
  {author} {\bibfnamefont {A.}~\bibnamefont {Aspuru-Guzik}},\ }\href@noop {}
  {\bibfield  {journal} {\bibinfo  {journal} {Nature Photonics}\ }\textbf
  {\bibinfo {volume} {9}},\ \bibinfo {pages} {615} (\bibinfo {year}
  {2015})}\BibitemShut {NoStop}%
\bibitem [{\citenamefont {Hamilton}\ \emph {et~al.}(2017)\citenamefont
  {Hamilton}, \citenamefont {Kruse}, \citenamefont {Sansoni}, \citenamefont
  {Barkhofen}, \citenamefont {Silberhorn},\ and\ \citenamefont
  {Jex}}]{hamilton2017gaussian}%
  \BibitemOpen
  \bibfield  {author} {\bibinfo {author} {\bibfnamefont {C.~S.}\ \bibnamefont
  {Hamilton}}, \bibinfo {author} {\bibfnamefont {R.}~\bibnamefont {Kruse}},
  \bibinfo {author} {\bibfnamefont {L.}~\bibnamefont {Sansoni}}, \bibinfo
  {author} {\bibfnamefont {S.}~\bibnamefont {Barkhofen}}, \bibinfo {author}
  {\bibfnamefont {C.}~\bibnamefont {Silberhorn}}, \ and\ \bibinfo {author}
  {\bibfnamefont {I.}~\bibnamefont {Jex}},\ }\href@noop {} {\bibfield
  {journal} {\bibinfo  {journal} {Physical Review Letters}\ }\textbf {\bibinfo
  {volume} {119}},\ \bibinfo {pages} {170501} (\bibinfo {year}
  {2017})}\BibitemShut {NoStop}%
\bibitem [{\citenamefont {Weedbrook}\ \emph {et~al.}(2012)\citenamefont
  {Weedbrook}, \citenamefont {Pirandola}, \citenamefont
  {Garc{\'\i}a-Patr{\'o}n}, \citenamefont {Cerf}, \citenamefont {Ralph},
  \citenamefont {Shapiro},\ and\ \citenamefont
  {Lloyd}}]{weedbrook2012gaussian}%
  \BibitemOpen
  \bibfield  {author} {\bibinfo {author} {\bibfnamefont {C.}~\bibnamefont
  {Weedbrook}}, \bibinfo {author} {\bibfnamefont {S.}~\bibnamefont
  {Pirandola}}, \bibinfo {author} {\bibfnamefont {R.}~\bibnamefont
  {Garc{\'\i}a-Patr{\'o}n}}, \bibinfo {author} {\bibfnamefont {N.~J.}\
  \bibnamefont {Cerf}}, \bibinfo {author} {\bibfnamefont {T.~C.}\ \bibnamefont
  {Ralph}}, \bibinfo {author} {\bibfnamefont {J.~H.}\ \bibnamefont {Shapiro}},
  \ and\ \bibinfo {author} {\bibfnamefont {S.}~\bibnamefont {Lloyd}},\
  }\href@noop {} {\bibfield  {journal} {\bibinfo  {journal} {Reviews of Modern
  Physics}\ }\textbf {\bibinfo {volume} {84}},\ \bibinfo {pages} {621}
  (\bibinfo {year} {2012})}\BibitemShut {NoStop}%
\bibitem [{\citenamefont {Menicucci}(2014)}]{menicucci2014fault}%
  \BibitemOpen
  \bibfield  {author} {\bibinfo {author} {\bibfnamefont {N.~C.}\ \bibnamefont
  {Menicucci}},\ }\href@noop {} {\bibfield  {journal} {\bibinfo  {journal}
  {Physical Review Letters}\ }\textbf {\bibinfo {volume} {112}},\ \bibinfo
  {pages} {120504} (\bibinfo {year} {2014})}\BibitemShut {NoStop}%
\bibitem [{\citenamefont {Marsili}\ \emph {et~al.}(2013)\citenamefont
  {Marsili}, \citenamefont {Verma}, \citenamefont {Stern}, \citenamefont
  {Harrington}, \citenamefont {Lita}, \citenamefont {Gerrits}, \citenamefont
  {Vayshenker}, \citenamefont {Baek}, \citenamefont {Shaw}, \citenamefont
  {Mirin} \emph {et~al.}}]{marsili2013detecting}%
  \BibitemOpen
  \bibfield  {author} {\bibinfo {author} {\bibfnamefont {F.}~\bibnamefont
  {Marsili}}, \bibinfo {author} {\bibfnamefont {V.~B.}\ \bibnamefont {Verma}},
  \bibinfo {author} {\bibfnamefont {J.~A.}\ \bibnamefont {Stern}}, \bibinfo
  {author} {\bibfnamefont {S.}~\bibnamefont {Harrington}}, \bibinfo {author}
  {\bibfnamefont {A.~E.}\ \bibnamefont {Lita}}, \bibinfo {author}
  {\bibfnamefont {T.}~\bibnamefont {Gerrits}}, \bibinfo {author} {\bibfnamefont
  {I.}~\bibnamefont {Vayshenker}}, \bibinfo {author} {\bibfnamefont
  {B.}~\bibnamefont {Baek}}, \bibinfo {author} {\bibfnamefont {M.~D.}\
  \bibnamefont {Shaw}}, \bibinfo {author} {\bibfnamefont {R.~P.}\ \bibnamefont
  {Mirin}},  \emph {et~al.},\ }\href@noop {} {\bibfield  {journal} {\bibinfo
  {journal} {Nature Photonics}\ }\textbf {\bibinfo {volume} {7}},\ \bibinfo
  {pages} {210} (\bibinfo {year} {2013})}\BibitemShut {NoStop}%
\bibitem [{\citenamefont {Arrazola}\ and\ \citenamefont
  {Bromley}(2018)}]{arrazola2018using}%
  \BibitemOpen
  \bibfield  {author} {\bibinfo {author} {\bibfnamefont {J.~M.}\ \bibnamefont
  {Arrazola}}\ and\ \bibinfo {author} {\bibfnamefont {T.~R.}\ \bibnamefont
  {Bromley}},\ }\href@noop {} {\bibfield  {journal} {\bibinfo  {journal}
  {arXiv:1803.10730}\ } (\bibinfo {year} {2018})}\BibitemShut {NoStop}%
\bibitem [{\citenamefont {Bradler}\ \emph {et~al.}(2017)\citenamefont
  {Bradler}, \citenamefont {Dallaire-Demers}, \citenamefont {Rebentrost},
  \citenamefont {Su},\ and\ \citenamefont {Weedbrook}}]{bradler2017gaussian}%
  \BibitemOpen
  \bibfield  {author} {\bibinfo {author} {\bibfnamefont {K.}~\bibnamefont
  {Bradler}}, \bibinfo {author} {\bibfnamefont {P.-L.}\ \bibnamefont
  {Dallaire-Demers}}, \bibinfo {author} {\bibfnamefont {P.}~\bibnamefont
  {Rebentrost}}, \bibinfo {author} {\bibfnamefont {D.}~\bibnamefont {Su}}, \
  and\ \bibinfo {author} {\bibfnamefont {C.}~\bibnamefont {Weedbrook}},\
  }\href@noop {} {\bibfield  {journal} {\bibinfo  {journal} {arXiv:1712.06729}\
  } (\bibinfo {year} {2017})}\BibitemShut {NoStop}%
\bibitem [{\citenamefont {Ramelow}\ \emph {et~al.}(2018)\citenamefont
  {Ramelow}, \citenamefont {Farsi}, \citenamefont {Vernon}, \citenamefont
  {Clemmen}, \citenamefont {Ji}, \citenamefont {Sipe}, \citenamefont
  {Liscidini}, \citenamefont {Lipson},\ and\ \citenamefont
  {Gaeta}}]{ramelow2018strong}%
  \BibitemOpen
  \bibfield  {author} {\bibinfo {author} {\bibfnamefont {S.}~\bibnamefont
  {Ramelow}}, \bibinfo {author} {\bibfnamefont {A.}~\bibnamefont {Farsi}},
  \bibinfo {author} {\bibfnamefont {Z.}~\bibnamefont {Vernon}}, \bibinfo
  {author} {\bibfnamefont {S.}~\bibnamefont {Clemmen}}, \bibinfo {author}
  {\bibfnamefont {X.}~\bibnamefont {Ji}}, \bibinfo {author} {\bibfnamefont
  {J.~E.}\ \bibnamefont {Sipe}}, \bibinfo {author} {\bibfnamefont
  {M.}~\bibnamefont {Liscidini}}, \bibinfo {author} {\bibfnamefont
  {M.}~\bibnamefont {Lipson}}, \ and\ \bibinfo {author} {\bibfnamefont {A.~L.}\
  \bibnamefont {Gaeta}},\ }\href@noop {} {\bibfield  {journal} {\bibinfo
  {journal} {arXiv:1802.10072}\ } (\bibinfo {year} {2018})}\BibitemShut
  {NoStop}%
\bibitem [{\citenamefont {Vernon}\ and\ \citenamefont
  {Sipe}(2015)}]{vernon2015spontaneous}%
  \BibitemOpen
  \bibfield  {author} {\bibinfo {author} {\bibfnamefont {Z.}~\bibnamefont
  {Vernon}}\ and\ \bibinfo {author} {\bibfnamefont {J.}~\bibnamefont {Sipe}},\
  }\href@noop {} {\bibfield  {journal} {\bibinfo  {journal} {Physical Review
  A}\ }\textbf {\bibinfo {volume} {91}},\ \bibinfo {pages} {053802} (\bibinfo
  {year} {2015})}\BibitemShut {NoStop}%
\bibitem [{\citenamefont {Hoff}\ \emph {et~al.}(2015)\citenamefont {Hoff},
  \citenamefont {Nielsen},\ and\ \citenamefont
  {Andersen}}]{hoff2015integrated}%
  \BibitemOpen
  \bibfield  {author} {\bibinfo {author} {\bibfnamefont {U.~B.}\ \bibnamefont
  {Hoff}}, \bibinfo {author} {\bibfnamefont {B.~M.}\ \bibnamefont {Nielsen}}, \
  and\ \bibinfo {author} {\bibfnamefont {U.~L.}\ \bibnamefont {Andersen}},\
  }\href@noop {} {\bibfield  {journal} {\bibinfo  {journal} {Optics express}\
  }\textbf {\bibinfo {volume} {23}},\ \bibinfo {pages} {12013} (\bibinfo {year}
  {2015})}\BibitemShut {NoStop}%
\bibitem [{\citenamefont {Okawachi}\ \emph {et~al.}(2015)\citenamefont
  {Okawachi}, \citenamefont {Yu}, \citenamefont {Luke}, \citenamefont
  {Carvalho}, \citenamefont {Ramelow}, \citenamefont {Farsi}, \citenamefont
  {Lipson},\ and\ \citenamefont {Gaeta}}]{okawachi2015dual}%
  \BibitemOpen
  \bibfield  {author} {\bibinfo {author} {\bibfnamefont {Y.}~\bibnamefont
  {Okawachi}}, \bibinfo {author} {\bibfnamefont {M.}~\bibnamefont {Yu}},
  \bibinfo {author} {\bibfnamefont {K.}~\bibnamefont {Luke}}, \bibinfo {author}
  {\bibfnamefont {D.~O.}\ \bibnamefont {Carvalho}}, \bibinfo {author}
  {\bibfnamefont {S.}~\bibnamefont {Ramelow}}, \bibinfo {author} {\bibfnamefont
  {A.}~\bibnamefont {Farsi}}, \bibinfo {author} {\bibfnamefont
  {M.}~\bibnamefont {Lipson}}, \ and\ \bibinfo {author} {\bibfnamefont {A.~L.}\
  \bibnamefont {Gaeta}},\ }\href@noop {} {\bibfield  {journal} {\bibinfo
  {journal} {Optics Letters}\ }\textbf {\bibinfo {volume} {40}},\ \bibinfo
  {pages} {5267} (\bibinfo {year} {2015})}\BibitemShut {NoStop}%
\bibitem [{\citenamefont {Reimer}\ \emph {et~al.}(2015)\citenamefont {Reimer},
  \citenamefont {Kues}, \citenamefont {Caspani}, \citenamefont {Wetzel},
  \citenamefont {Roztocki}, \citenamefont {Clerici}, \citenamefont {Jestin},
  \citenamefont {Ferrera}, \citenamefont {Peccianti}, \citenamefont {Pasquazi}
  \emph {et~al.}}]{reimer2015cross}%
  \BibitemOpen
  \bibfield  {author} {\bibinfo {author} {\bibfnamefont {C.}~\bibnamefont
  {Reimer}}, \bibinfo {author} {\bibfnamefont {M.}~\bibnamefont {Kues}},
  \bibinfo {author} {\bibfnamefont {L.}~\bibnamefont {Caspani}}, \bibinfo
  {author} {\bibfnamefont {B.}~\bibnamefont {Wetzel}}, \bibinfo {author}
  {\bibfnamefont {P.}~\bibnamefont {Roztocki}}, \bibinfo {author}
  {\bibfnamefont {M.}~\bibnamefont {Clerici}}, \bibinfo {author} {\bibfnamefont
  {Y.}~\bibnamefont {Jestin}}, \bibinfo {author} {\bibfnamefont
  {M.}~\bibnamefont {Ferrera}}, \bibinfo {author} {\bibfnamefont
  {M.}~\bibnamefont {Peccianti}}, \bibinfo {author} {\bibfnamefont
  {A.}~\bibnamefont {Pasquazi}},  \emph {et~al.},\ }\href@noop {} {\bibfield
  {journal} {\bibinfo  {journal} {Nature Communications}\ }\textbf {\bibinfo
  {volume} {6}},\ \bibinfo {pages} {8236} (\bibinfo {year} {2015})}\BibitemShut
  {NoStop}%
\bibitem [{\citenamefont {He}\ \emph {et~al.}(2015)\citenamefont {He},
  \citenamefont {Bell}, \citenamefont {Casas-Bedoya}, \citenamefont {Zhang},
  \citenamefont {Clark}, \citenamefont {Xiong},\ and\ \citenamefont
  {Eggleton}}]{he2015ultracompact}%
  \BibitemOpen
  \bibfield  {author} {\bibinfo {author} {\bibfnamefont {J.}~\bibnamefont
  {He}}, \bibinfo {author} {\bibfnamefont {B.~A.}\ \bibnamefont {Bell}},
  \bibinfo {author} {\bibfnamefont {A.}~\bibnamefont {Casas-Bedoya}}, \bibinfo
  {author} {\bibfnamefont {Y.}~\bibnamefont {Zhang}}, \bibinfo {author}
  {\bibfnamefont {A.~S.}\ \bibnamefont {Clark}}, \bibinfo {author}
  {\bibfnamefont {C.}~\bibnamefont {Xiong}}, \ and\ \bibinfo {author}
  {\bibfnamefont {B.~J.}\ \bibnamefont {Eggleton}},\ }\href@noop {} {\bibfield
  {journal} {\bibinfo  {journal} {Optica}\ }\textbf {\bibinfo {volume} {2}},\
  \bibinfo {pages} {779} (\bibinfo {year} {2015})}\BibitemShut {NoStop}%
\bibitem [{\citenamefont {Guo}\ \emph {et~al.}(2014)\citenamefont {Guo},
  \citenamefont {Zhang}, \citenamefont {Dong}, \citenamefont {Huang},\ and\
  \citenamefont {Peng}}]{guo2014telecom}%
  \BibitemOpen
  \bibfield  {author} {\bibinfo {author} {\bibfnamefont {Y.}~\bibnamefont
  {Guo}}, \bibinfo {author} {\bibfnamefont {W.}~\bibnamefont {Zhang}}, \bibinfo
  {author} {\bibfnamefont {S.}~\bibnamefont {Dong}}, \bibinfo {author}
  {\bibfnamefont {Y.}~\bibnamefont {Huang}}, \ and\ \bibinfo {author}
  {\bibfnamefont {J.}~\bibnamefont {Peng}},\ }\href@noop {} {\bibfield
  {journal} {\bibinfo  {journal} {Optics Letters}\ }\textbf {\bibinfo {volume}
  {39}},\ \bibinfo {pages} {2526} (\bibinfo {year} {2014})}\BibitemShut
  {NoStop}%
\bibitem [{\citenamefont {Vernon}\ \emph {et~al.}(2016)\citenamefont {Vernon},
  \citenamefont {Liscidini},\ and\ \citenamefont {Sipe}}]{vernon2016quantum}%
  \BibitemOpen
  \bibfield  {author} {\bibinfo {author} {\bibfnamefont {Z.}~\bibnamefont
  {Vernon}}, \bibinfo {author} {\bibfnamefont {M.}~\bibnamefont {Liscidini}}, \
  and\ \bibinfo {author} {\bibfnamefont {J.}~\bibnamefont {Sipe}},\ }\href@noop
  {} {\bibfield  {journal} {\bibinfo  {journal} {Physical Review A}\ }\textbf
  {\bibinfo {volume} {94}},\ \bibinfo {pages} {023810} (\bibinfo {year}
  {2016})}\BibitemShut {NoStop}%
\bibitem [{\citenamefont {Li}\ \emph {et~al.}(2016)\citenamefont {Li},
  \citenamefont {Davan{\c{c}}o},\ and\ \citenamefont
  {Srinivasan}}]{li2016efficient}%
  \BibitemOpen
  \bibfield  {author} {\bibinfo {author} {\bibfnamefont {Q.}~\bibnamefont
  {Li}}, \bibinfo {author} {\bibfnamefont {M.}~\bibnamefont {Davan{\c{c}}o}}, \
  and\ \bibinfo {author} {\bibfnamefont {K.}~\bibnamefont {Srinivasan}},\
  }\href@noop {} {\bibfield  {journal} {\bibinfo  {journal} {Nature Photonics}\
  }\textbf {\bibinfo {volume} {10}},\ \bibinfo {pages} {406} (\bibinfo {year}
  {2016})}\BibitemShut {NoStop}%
\bibitem [{\citenamefont {Gentry}\ \emph {et~al.}(2014)\citenamefont {Gentry},
  \citenamefont {Zeng},\ and\ \citenamefont {Popovi{\'c}}}]{gentry2014tunable}%
  \BibitemOpen
  \bibfield  {author} {\bibinfo {author} {\bibfnamefont {C.~M.}\ \bibnamefont
  {Gentry}}, \bibinfo {author} {\bibfnamefont {X.}~\bibnamefont {Zeng}}, \ and\
  \bibinfo {author} {\bibfnamefont {M.~A.}\ \bibnamefont {Popovi{\'c}}},\
  }\href@noop {} {\bibfield  {journal} {\bibinfo  {journal} {Optics Letters}\
  }\textbf {\bibinfo {volume} {39}},\ \bibinfo {pages} {5689} (\bibinfo {year}
  {2014})}\BibitemShut {NoStop}%
\bibitem [{\citenamefont {Chen}\ \emph {et~al.}(2007)\citenamefont {Chen},
  \citenamefont {Sherwood-Droz},\ and\ \citenamefont
  {Lipson}}]{chen2007compact}%
  \BibitemOpen
  \bibfield  {author} {\bibinfo {author} {\bibfnamefont {L.}~\bibnamefont
  {Chen}}, \bibinfo {author} {\bibfnamefont {N.}~\bibnamefont {Sherwood-Droz}},
  \ and\ \bibinfo {author} {\bibfnamefont {M.}~\bibnamefont {Lipson}},\
  }\href@noop {} {\bibfield  {journal} {\bibinfo  {journal} {Optics Letters}\
  }\textbf {\bibinfo {volume} {32}},\ \bibinfo {pages} {3361} (\bibinfo {year}
  {2007})}\BibitemShut {NoStop}%
\bibitem [{\citenamefont {Moss}\ \emph {et~al.}(2013)\citenamefont {Moss},
  \citenamefont {Morandotti}, \citenamefont {Gaeta},\ and\ \citenamefont
  {Lipson}}]{moss2013new}%
  \BibitemOpen
  \bibfield  {author} {\bibinfo {author} {\bibfnamefont {D.~J.}\ \bibnamefont
  {Moss}}, \bibinfo {author} {\bibfnamefont {R.}~\bibnamefont {Morandotti}},
  \bibinfo {author} {\bibfnamefont {A.~L.}\ \bibnamefont {Gaeta}}, \ and\
  \bibinfo {author} {\bibfnamefont {M.}~\bibnamefont {Lipson}},\ }\href@noop {}
  {\bibfield  {journal} {\bibinfo  {journal} {Nature Photonics}\ }\textbf
  {\bibinfo {volume} {7}},\ \bibinfo {pages} {597} (\bibinfo {year}
  {2013})}\BibitemShut {NoStop}%
\bibitem [{\citenamefont {Vernon}\ \emph {et~al.}(2017)\citenamefont {Vernon},
  \citenamefont {Menotti}, \citenamefont {Tison}, \citenamefont {Steidle},
  \citenamefont {Fanto}, \citenamefont {Thomas}, \citenamefont {Preble},
  \citenamefont {Smith}, \citenamefont {Alsing}, \citenamefont {Liscidini}
  \emph {et~al.}}]{vernon2017truly}%
  \BibitemOpen
  \bibfield  {author} {\bibinfo {author} {\bibfnamefont {Z.}~\bibnamefont
  {Vernon}}, \bibinfo {author} {\bibfnamefont {M.}~\bibnamefont {Menotti}},
  \bibinfo {author} {\bibfnamefont {C.}~\bibnamefont {Tison}}, \bibinfo
  {author} {\bibfnamefont {J.}~\bibnamefont {Steidle}}, \bibinfo {author}
  {\bibfnamefont {M.}~\bibnamefont {Fanto}}, \bibinfo {author} {\bibfnamefont
  {P.}~\bibnamefont {Thomas}}, \bibinfo {author} {\bibfnamefont
  {S.}~\bibnamefont {Preble}}, \bibinfo {author} {\bibfnamefont
  {A.}~\bibnamefont {Smith}}, \bibinfo {author} {\bibfnamefont
  {P.}~\bibnamefont {Alsing}}, \bibinfo {author} {\bibfnamefont
  {M.}~\bibnamefont {Liscidini}},  \emph {et~al.},\ }\href@noop {} {\bibfield
  {journal} {\bibinfo  {journal} {Optics Letters}\ }\textbf {\bibinfo {volume}
  {42}},\ \bibinfo {pages} {3638} (\bibinfo {year} {2017})}\BibitemShut
  {NoStop}%
\bibitem [{\citenamefont {Christensen}\ \emph {et~al.}(2018)\citenamefont
  {Christensen}, \citenamefont {Koefoed}, \citenamefont {Rottwitt},\ and\
  \citenamefont {McKinstrie}}]{christensen2018engineering}%
  \BibitemOpen
  \bibfield  {author} {\bibinfo {author} {\bibfnamefont {J.}~\bibnamefont
  {Christensen}}, \bibinfo {author} {\bibfnamefont {J.}~\bibnamefont
  {Koefoed}}, \bibinfo {author} {\bibfnamefont {K.}~\bibnamefont {Rottwitt}}, \
  and\ \bibinfo {author} {\bibfnamefont {C.}~\bibnamefont {McKinstrie}},\
  }\href@noop {} {\bibfield  {journal} {\bibinfo  {journal} {Optics Letters}\
  }\textbf {\bibinfo {volume} {43}},\ \bibinfo {pages} {859} (\bibinfo {year}
  {2018})}\BibitemShut {NoStop}%
\bibitem [{\citenamefont {Tison}\ \emph {et~al.}(2017)\citenamefont {Tison},
  \citenamefont {Steidle}, \citenamefont {Fanto}, \citenamefont {Wang},
  \citenamefont {Mogent}, \citenamefont {Rizzo}, \citenamefont {Preble},\ and\
  \citenamefont {Alsing}}]{tison2017path}%
  \BibitemOpen
  \bibfield  {author} {\bibinfo {author} {\bibfnamefont {C.}~\bibnamefont
  {Tison}}, \bibinfo {author} {\bibfnamefont {J.}~\bibnamefont {Steidle}},
  \bibinfo {author} {\bibfnamefont {M.}~\bibnamefont {Fanto}}, \bibinfo
  {author} {\bibfnamefont {Z.}~\bibnamefont {Wang}}, \bibinfo {author}
  {\bibfnamefont {N.}~\bibnamefont {Mogent}}, \bibinfo {author} {\bibfnamefont
  {A.}~\bibnamefont {Rizzo}}, \bibinfo {author} {\bibfnamefont
  {S.}~\bibnamefont {Preble}}, \ and\ \bibinfo {author} {\bibfnamefont
  {P.}~\bibnamefont {Alsing}},\ }\href@noop {} {\bibfield  {journal} {\bibinfo
  {journal} {Optics Express}\ }\textbf {\bibinfo {volume} {25}},\ \bibinfo
  {pages} {33088} (\bibinfo {year} {2017})}\BibitemShut {NoStop}%
\bibitem [{\citenamefont {Braunstein}\ and\ \citenamefont
  {Kimble}(1998)}]{braunstein1998teleportation}%
  \BibitemOpen
  \bibfield  {author} {\bibinfo {author} {\bibfnamefont {S.~L.}\ \bibnamefont
  {Braunstein}}\ and\ \bibinfo {author} {\bibfnamefont {H.}~\bibnamefont
  {Kimble}},\ }\href@noop {} {\bibfield  {journal} {\bibinfo  {journal}
  {Physical Review Letters}\ }\textbf {\bibinfo {volume} {80}},\ \bibinfo
  {pages} {869} (\bibinfo {year} {1998})}\BibitemShut {NoStop}%
\bibitem [{\citenamefont {Spencer}\ \emph {et~al.}(2014)\citenamefont
  {Spencer}, \citenamefont {Bauters}, \citenamefont {Heck},\ and\ \citenamefont
  {Bowers}}]{spencer2014integrated}%
  \BibitemOpen
  \bibfield  {author} {\bibinfo {author} {\bibfnamefont {D.~T.}\ \bibnamefont
  {Spencer}}, \bibinfo {author} {\bibfnamefont {J.~F.}\ \bibnamefont
  {Bauters}}, \bibinfo {author} {\bibfnamefont {M.~J.}\ \bibnamefont {Heck}}, \
  and\ \bibinfo {author} {\bibfnamefont {J.~E.}\ \bibnamefont {Bowers}},\
  }\href@noop {} {\bibfield  {journal} {\bibinfo  {journal} {Optica}\ }\textbf
  {\bibinfo {volume} {1}},\ \bibinfo {pages} {153} (\bibinfo {year}
  {2014})}\BibitemShut {NoStop}%
\bibitem [{\citenamefont {Serafini}(2017)}]{serafini2017quantum}%
  \BibitemOpen
  \bibfield  {author} {\bibinfo {author} {\bibfnamefont {A.}~\bibnamefont
  {Serafini}},\ }\href@noop {} {\emph {\bibinfo {title} {Quantum Continuous
  Variables: A Primer of Theoretical Methods}}}\ (\bibinfo  {publisher} {CRC
  Press},\ \bibinfo {year} {2017})\BibitemShut {NoStop}%
\bibitem [{\citenamefont {Caves}(2017)}]{caves2017}%
  \BibitemOpen
  \bibfield  {author} {\bibinfo {author} {\bibfnamefont {C.}~\bibnamefont
  {Caves}},\ }\href
  {http://info.phys.unm.edu/~caves/courses/qinfo-s17/lectures/polarsingularAutonne.pdf}
  {\enquote {\bibinfo {title} {Polar decomposition, singular-value
  decomposition, and autonne-takagi factorization},}\ }\bibinfo {howpublished}
  {Quantum Information Lecture Notes} (\bibinfo {year} {2017})\BibitemShut
  {NoStop}%
\bibitem [{\citenamefont {Horn}\ and\ \citenamefont
  {Johnson}(1990)}]{horn1990matrix}%
  \BibitemOpen
  \bibfield  {author} {\bibinfo {author} {\bibfnamefont {R.~A.}\ \bibnamefont
  {Horn}}\ and\ \bibinfo {author} {\bibfnamefont {C.~R.}\ \bibnamefont
  {Johnson}},\ }\href@noop {} {\emph {\bibinfo {title} {Matrix analysis}}}\
  (\bibinfo  {publisher} {Cambridge university press},\ \bibinfo {year}
  {1990})\BibitemShut {NoStop}%
\bibitem [{\citenamefont {Killoran}\ \emph {et~al.}(2018)\citenamefont
  {Killoran}, \citenamefont {Izaac}, \citenamefont {Quesada}, \citenamefont
  {Bergholm}, \citenamefont {Amy},\ and\ \citenamefont
  {Weedbrook}}]{killoran2018strawberry}%
  \BibitemOpen
  \bibfield  {author} {\bibinfo {author} {\bibfnamefont {N.}~\bibnamefont
  {Killoran}}, \bibinfo {author} {\bibfnamefont {J.}~\bibnamefont {Izaac}},
  \bibinfo {author} {\bibfnamefont {N.}~\bibnamefont {Quesada}}, \bibinfo
  {author} {\bibfnamefont {V.}~\bibnamefont {Bergholm}}, \bibinfo {author}
  {\bibfnamefont {M.}~\bibnamefont {Amy}}, \ and\ \bibinfo {author}
  {\bibfnamefont {C.}~\bibnamefont {Weedbrook}},\ }\href@noop {} {\bibfield
  {journal} {\bibinfo  {journal} {arXiv preprint arXiv:1804.03159}\ } (\bibinfo
  {year} {2018})}\BibitemShut {NoStop}%
\bibitem [{\citenamefont {Quesada}(2015)}]{quesada2015very}%
  \BibitemOpen
  \bibfield  {author} {\bibinfo {author} {\bibfnamefont {N.}~\bibnamefont
  {Quesada}},\ }\emph {\bibinfo {title} {Very Nonlinear Quantum Optics}},\
  \href@noop {} {Ph.D. thesis},\ \bibinfo  {school} {University of Toronto
  (Canada)} (\bibinfo {year} {2015})\BibitemShut {NoStop}%
\end{thebibliography}%

\pagebreak
\widetext
\begin{center}
\textbf{\large Supplemental Information for \emph{Scalable squeezed light source for continuous variable quantum sampling}}
\end{center}
\setcounter{equation}{0}
\setcounter{figure}{0}
\setcounter{table}{0}
\setcounter{page}{1}
\makeatletter
\renewcommand{\theequation}{S\arabic{equation}}
\renewcommand{\thefigure}{S\arabic{figure}}

\section{Unwanted nonlinear effects}
A typical microresonator system accommodates many hundreds or even thousands of resonances. The term $H_X$ in Eq. 1 of the main text contains the corresponding contributions to the Hamiltonian, and their couplings to the three modes of interest. Below any thresholds for OPO and comb generation, and operating in a regime where cascaded four-wave mixing is negligible, there are two dominant unwanted couplings relevant to the device performance: those that give rise to unwanted spontaneous four-wave mixing, leading to the generation of spurious photons in the $S$ mode \cite{vernon2015spontaneous}, and those that give rise to Bragg-scattering four-wave mixing (BS-FWM), leading to an additional source of loss on the squeezed state generated in the $S$ mode \cite{vernon2016quantum}. These effects arise from terms of the form $b_Db_Db_S^\dagger b_{X1}^\dagger$ and $b_Pb_Pb_S^\dagger b_{X2}^\dagger$ (unwanted SFWM), and $b_D b_P^\dagger b_S b_{X1}^\dagger$ and $b_D^\dagger b_P b_S b_{X2}^\dagger$ (unwanted BS-FWM),  where $b_{X1}$ and $b_{X2}$ are annihilation operators for unwanted modes $X1$ and $X2$. 

Such processes effectively entangle the $X1$ and $X2$ modes with the $S$ mode, corrupting the purity of the $S$ mode output state. Though both normal dispersion and the effects of SPM/XPM from the strong drive act to counteract this effect by increasing the corresponding net detuning for these processes, in general for simple single-resonator devices with realistic parameters this is not sufficient to suppress the unwanted processes to a level appropriate for CV quantum sampling, in which highly pure Gaussian states are desirable. 

To see this first we estimate the strength of the SFWM process associated
with the generation of photons at $\omega_{S}$ and $\omega_{X2}$.
Similar consideration can be done for the pump field and pair generation
at $\omega_{S}$ and $\omega_{X1}$. However, the noise arising from
the driving field is expected to be several orders of magnitude larger,
for SFWM scales quadratically with the power of the generating field. 

The intensity of SFWM involving the modes at $\omega_{X2}$, $\omega_{S}$
and $\omega_{D}$ depends on the relative position of the corresponding
resonances. In this case, SPM and XPM leads to a relative detuning:

\begin{equation}
\delta=(\omega_{D}-\omega_{X_{2}})-(\omega_{S}-\omega_{D})=-\frac{3c}{\omega_{D}n_\mathrm{eff}}\gamma_\mathrm{NL}\frac{v_{g}Q}{2\pi R}P_{D},\label{eq:detuning}
\end{equation}

where $n_\mathrm{eff}$ is the mode effective index, $\gamma_\mathrm{NL}$ the waveguide
nonlinear parameter, and $P_{D}$ the input power at $\omega_{D}$.
Finally, $v_{g}$ is the group velocity, $Q$ the resonator quality
factor, and $R$ the ring radius.

In the presence of a detuning $\delta$, and in the limit of small
pair generation rate, the average number of generated pairs is reduced
according to

\[
|\beta_{\delta}|^{2}=|\beta_{0}|^{2}\frac{\Delta^{2}}{\delta^{2}+\Delta^{2}},
\]

where $|\beta_{0}|^{2}$ is the average number of pairs in the case
of equally spaced resonances (i.e. $\delta=0$) and $\Delta$ is the
resonance line width, which for simplicity we assume to be the same
for all the three resonances involved in the process.

Considering only the noise contribution coming from unwanted SFWM,
the signal-to-noise ratio (SNR) can be defied as

\[
SNR=\frac{|\beta_{\mathrm{S}}|^{2}}{|\beta_{\mathrm{S,X_{2}}}|^{2}},
\]

where $|\beta_{\mathrm{S}}|^{2}$ and $|\beta_{\mathrm{S,X_{2}}}|^{2}$
are the average number of pairs generated in the signal mode and by
the unwanted SFWM associated to the driving field respectively. This
leads to

\begin{equation}
SNR=\frac{|\beta_{\mathrm{S}}|^{2}}{|\beta_{\mathrm{S,X_{2}}}|^{2}}\approx\frac{P_{P}}{P_{D}}\left(1+\frac{\delta^{2}}{\Delta^{2}}\right),\label{eq:SNR_2}
\end{equation}

where $P_{P}$ is the pump field power. As expected, in the absence
of detuning, i.e. when all the resonances at $\omega_{X_{2}}$, $\omega_{D}$,
$\omega_{S}$, and $\omega_{P}$ are equally spaced, the SNR would
be essentially proportional to $P_{P}/P_{D}$. However, due to the
presence of SPM/XPM, the unwanted SFWM process is suppressed by the
detuning. We can re write (\ref{eq:SNR_2}) by explicitly taking into
account the dependence on the structure parameters using (\ref{eq:detuning}):

\begin{equation}
SNR=\approx\frac{P_{P}}{P_{D}}\left(1+\xi^{2}\frac{Q^{4}}{R^{2}}P_{D}^{2}\right),\label{eq:SNR_structure_paramters}
\end{equation}

where 

\begin{equation}
\xi=\frac{3c^{2}\gamma_\mathrm{NL}}{2\pi\omega^{2}n_\mathrm{eff}n_{g}}\label{eq:csi_parameter}
\end{equation}

is a constant that depends on the nonlinearity and dispersion of the
waveguide. 

In the case of SiN ring resonators, one can take $\xi=10^{-14}\mathrm{m}\cdot\mathrm{W}^{-1}$,
$Q=10^{6},$ $R=10^{-4}\,\mathrm{m,}$ $P_{D}=2\cdot10^{-1}W,$ and
$P_{P}=10^{-3}\,W$, which lead to

\[
SNR\approx2.
\]

This value suggests that, in general, the resonance detuning determined
by SPM/XPM is not sufficient to neglect the effect of unwanted pairs
generated by the driving field via degenerate SFWM.

Care must therefore be taken to further corrupt the efficiency of the unwanted process, either by detuning or removing altogether the associated resonances $X1$ and $X2$. Many strategies exist to accomplish this without significantly compromising other aspects of device performance: one particularly promising possibility is to couple to the primary resonator two auxiliary resonators with different free spectral ranges, which can be used to selectively split and severely detune the $X1$ and $X2$ resonances from their default frequencies \cite{gentry2014tunable}. Another strategy involves using a Mach-Zehnder interferometer-based coupler to independently modify the quality factors of the resonances; the efficiency of processes involving the unwanted $X1$ and $X2$ modes can be strongly degraded by reducing their associated quality factors. Alternatively, if a more sophisticated coupled resonator system is used to obviate the need for strong dispersion, and the associated free spectral ranges are chosen to be incommensurate, the unwanted resonances will be absent, strongly suppressing unwanted four-wave mixing effects.

\section{Temporal mode decomposition}
In this section we analyze how, starting from the second order moments of the channel fields, one can obtain the quantum state of systems that produce zero-mean Gaussian states, and calculate any measureable quantity related to the device output. We start by noting that the intra-resonator dynamics are \emph{linear in the quantum operators} within our assumptions, and thus since the initial quantum state is vacuum the state at all times is Gaussian\cite{serafini2017quantum}, i.e, it is described by a mean displacement 
\begin{align}
\braket{\psi_{S,\text{out}}(t)},
\end{align}
and the two point correlation functions
\begin{align}
N(t,t') = v_S \braket{\psi^\dagger_{S,\text{out}}(t) \psi_{S,\text{out}}(t')},\\
M(t,t') = v_S \braket{\psi_{S,\text{out}}(t) \psi_{S,\text{out}}(t')}.
\end{align}
For our system we can always guarantee that $\braket{\psi_{S,\mathrm{out}}(t)}=0$ and thus for the rest of this section we focus on $M(t,t')$ and $N(t,t')$.
Furthermote note that $N$ is hermitian and $M$ is symmetric
\begin{align}
N(t,t') = N(t',t)^* \text{ and } M(t,t') = M(t',t).
\end{align}
In the absence of intrinsic losses, i.e., assuming that any photon created inside the resonator can only leak into the waveguide at rate $\Gamma_S$, we know that the quantum state of the waveguide is pure once the resonator modes populations have decayed. If this were not the case there would be some entanglement between the mode $S$ in the resonator and the mode $S$ in the waveguide and thus the resonator could not be in the vacuum state. 
Knowing this we can use Williamson's theorem and the Bloch-Messiah decomposition \cite{serafini2017quantum} to write a joint decomposition of the second order moments as follows
\begin{align}\label{decomp}
N(t,t';\Gamma_S)_{\text{pure}} &= \sum_{\lambda } \sinh(r_\lambda)^2 f_\lambda(t) f^*_\lambda(t'),\\
M(t,t';\Gamma_S)_{\text{pure}} &= \sum_{\lambda } \frac{\sinh(2 r_\lambda)}{2} f^*_\lambda(t) f^*_\lambda(t'),
\end{align}
where the set of functions $f_{\lambda}(t)$ are orthonormal and complete
\begin{align}
\int dt \  f_{\lambda}(t) f^*_{\lambda'}(t) &= \delta_{\lambda,\lambda'}, \\
\sum_{\lambda} f_{\lambda}(t) f^*_{\lambda}(t') &= \delta(t-t').
\end{align}
and we use the subindex $\text{pure}$ to indicate that these moments come from a pure state and we explicitly write the dependence on the overall decay rate $\Gamma_S$ into the waveguide.
For pure states there is a certain degree of redundancy since if one has just the $N$ moment one can obtain the set $\{f_{\lambda}(t) \}$ and the mean photon numbers $\sinh^2( r_\lambda)$ via a simple eigendecomposition. Alternatively, if one has the $M$ moment one can obtain the set $\{f_{\lambda}(t) \}$ and the quantities $\sinh( 2 r_\lambda)/2$ via a Takagi-Autonne decomposition \cite{caves2017,horn1990matrix}. For our purposes we used the eigendecomposition of $N$ but also verified consistency using the Takagi-Autonne decomposition as implemented in Strawberry Fields \cite{killoran2018strawberry}. Note that in practice one knows the correlators in a grid of points and then for a sufficiently dense grid estimates the decompositions in Eq. (\ref{decomp}) (c.f. Appendix B of Ref.  \cite{quesada2015very}). Having the functions $f_\lambda(t)$ and the coefficients $r_\lambda$ one can write the ket describing the state of the waveguide as
\begin{align}
\ket{\Psi}  &= \bigotimes_{\lambda} S(r_\lambda, A_\lambda) \ket{\text{vac}},\\
 S(r_\lambda, A_\lambda)&=\exp\left(\frac{r_\lambda}{2} \left[A_\lambda^2 - A_\lambda^{\dagger 2} \right] \right),\\
A_\lambda & = \int dt \  \psi_{S,out}(t) f^*_\lambda(t).
\end{align}
Now let us conside the case where photons from the ring can be scattered into modes different from the ones in the waveguide, e.g. scattering modes that contribute to loss of photons from the resonator. The treatment of such losses into an undesired channel has been dealt with in e.g. \cite{vernon2015spontaneous}. The system is now decribed by several decay rates. The first one is $\Gamma_S$ the decay rate into the waveguide, then there is $M_S$ the scattering loss decay rate and finally there is $\overline{\Gamma}_S = \Gamma_S+M_S$ which is the total decay rate. For our correlation functions it is easily seen that in the case of loss the moments associated with this mixed state are related to the moments of a pure state as in Eq. (\ref{decomp}) where all the photons go into a fictitious waveguide at rate $\overline{\Gamma}_S$, as follows
\begin{align}\label{mixed}
N_{\text{mixed}}(t,t') &= \eta^{\text{esc}}_S N_{\text{pure}}(t,t';\bar \Gamma_S)\\
M_{\text{mixed}}(t,t') &= \eta^{\text{esc}}_S M_{\text{pure}}(t,t';\bar \Gamma_S)
\end{align}
where $\eta^{\text{esc}}_S = \Gamma_S/{\overline{\Gamma}_S}$ is the escape efficiency into the channel, i.e., the ratio of the decay rate into the physical waveguide to the overall decay rate into all channels (inclusing the waveguide). If there is only decay into the waveguide we have $\eta_{\text{esc}}=1, \ \Gamma_S = \bar \Gamma_S$ and recover the pure state case discussed previously.

The moments in Eq. (\ref{mixed}) are also the moments of the state $\ket{\Psi}$ after being sent through a loss channel where the (energy) tranmission is precisely $\eta_S^{\text{esc}}$. Note that a squeezed state with squeezing parameter $r$ subjected to loss by amount $\eta$ has the same density matrix as a thermal state with mean photon number 
\begin{align}\label{n}
\bar n = \eta \sinh^2 r
\end{align} 
and then squeezed by amount 
\begin{align}\label{np}
r' = \tanh^{-1} \left(  \frac{\eta \sinh(r) \cosh(r)}{1+\eta \sinh^2 r}\right).
\end{align}
Using this equivalence we write the density matrix of the state when scattering into unmeasured modes is present as
\begin{align}
\rho = \bigotimes_{\lambda}  S(r_\lambda' , A_\lambda) \left\{  \rho_{\lambda}(\bar n_\lambda) \right\} S(r_\lambda' , A_\lambda) ^\dagger.
\end{align}
where $\rho_\lambda$ is a thermal state in the mode with temporal profile $\lambda$ and with mean occupation number $\bar n_\lambda$ given by Eq. (\ref{n}) with $\eta = \eta_{\text{esc}}$ and $r = r_\lambda$. Likewise $r_\lambda'$ is given in Eq. (\ref{np}) with the same substitutions.

\end{document}